\documentclass[times,12pt,a4paper]{article}
\usepackage[left = 2.5cm, top = 2.5cm, right = 2.5cm, bottom = 2.5cm]{geometry}

\usepackage{amsmath}

\usepackage{pgf,pgfarrows,pgfnodes,pgfautomata,pgfheaps,pgfshade,bbm,bm,rotating,multirow,dsfont,bigdelim}
\usepackage{enumitem}
\usepackage{authblk}
\usepackage{graphicx}
\usepackage{bigints}
\usepackage{soul}
\usepackage{color}
\usepackage{tablefootnote}
\usepackage{xcolor}
\usepackage{listings}
\usepackage{amssymb}
\usepackage{multirow,bigdelim}
\usepackage{caption}
\usepackage{float}
\usepackage{tabularx}
\usepackage{MnSymbol}
\usepackage{subfigure}
\usepackage{makecell}
\usepackage{epstopdf}
\usepackage{fancyhdr,afterpage}
\usepackage{url}
\usepackage{enumitem}
\usepackage{amsmath,amssymb}
\usepackage[latin1]{inputenc}
\usepackage{colortbl}
\usepackage[english]{babel}
\usepackage{times}
\usepackage{makecell}
\usepackage{caption}
\usepackage{setspace}
\usepackage{graphicx}
\usepackage{textcomp}
\usepackage{makecell}
\usepackage{bbm,bm,rotating,multirow,dsfont,bigdelim}
\usepackage{perpage}
\MakePerPage{footnote}
\usepackage{textpos}
\usepackage{tikz}

\usetikzlibrary{arrows,positioning}
\tikzstyle{ov}=[shape=rectangle,\cite{}
                draw=black!50,
                line width= 1pt,
                minimum width=0.7cm,
                minimum height=0.7cm]

\tikzstyle{av}=[shape=rectangle,
                draw=black!50,
                fill=black!1,
                line width=1pt,
                minimum width=0.7cm,
                minimum height=0.7cm]

\tikzstyle{bv}=[shape=rectangle,
                draw=black!50,
                fill=black!1,
		   dashed,
                line width=1pt,
                minimum width=0.5cm,
                minimum height=0.5cm]

\tikzstyle{wv}=[shape=circle,draw=white,thick]
\tikzstyle{lv}=[shape=circle,draw=black!50,thick]
\tikzstyle{lvb}=[shape=circle,draw=black!0.50,thick]
\tikzstyle{lvr}=[shape=circle,draw=black]
\tikzstyle{svb}=[shape=rectangle,draw=black!0.50,thick]

\usepackage{graphicx}
\usepackage{epstopdf}

\usepackage{booktabs}
\usepackage{array}
\usepackage{verbatim}

\def\ci{\perp\!\!\!\perp}

\providecommand{\keywords}[1]{\textbf{\textit{Keywords:}} #1}
\title{Two Stage Least Squares with Time-Varying Instruments: An Application to an Evaluation of Treatment Intensification for Type-2 Diabetes}
\author[1]{Daniel Tompsett}
\author[2]{Stijn Vansteelandt}
\author[3]{Richard Grieve}
\author[4]{Irene Petersen}
\author[5]{Manuel Gomes}
\small{
\affil[1]{\tiny Dr Daniel Tompsett, Primary Care and Population Health, University College London, United Kingdom, email: d.tompsett@ucl.ac.uk}
\affil[2]{\tiny  Professor Stijn Vansteelandt, Department of Applied Mathematics, Computer Science, and Statistics, University of Ghent, Belgium, email: Stijn.Vansteelandt@UGent.be}
\affil[3]{\tiny Professor Richard Grieve, Centre for Data and Statistical Science for Health (DASH), London School of Hygiene and Tropical Medicine, United Kingdom, email: Richard.Grieve@lshtm.ac.uk}
\affil[4]{\tiny Professor Irene Petersen, Primary Care and Population Health, University College London, United Kingdom, email: i.petersen@ucl.ac.uk}
\affil[5]{\tiny Professor Manuel Gomes, Primary Care and Population Health, University College London, United Kingdom, email: m.gomes@ucl.ac.uk}}
\date{}
\begin{document}
\maketitle
\abstract
As longitudinal data becomes more available in many settings, policy makers are increasingly interested in the effect of time-varying treatments (e.g. sustained treatment strategies). In settings such as this, the preferred analysis techniques are the g-methods, however these require the untestable assumption of no unmeasured confounding. Instrumental variable analyses can minimise bias through unmeasured confounding. Of these methods, the Two Stage Least Squares technique is one of the most well used in Econometrics, but it has not been fully extended, and evaluated, in full time-varying settings. This paper proposes a robust two stage least squares method for the econometric evaluation of time-varying treatment. Using a simulation study we found that, unlike standard two stage least squares, it performs relatively well across a wide range of circumstances, including model misspecification. It compares well with recent time-varying instrument approaches via g-estimation. We illustrate the methods in an evaluation of treatment intensification for Type-2 Diabetes Mellitus, exploring the exogeneity in prescribing preferences to operationalise a time-varying instrument. 

\section{Introduction}
As routinely collected data has become more available, there has been an increasing interest in studies identifying time-varying treatment effects in observational databases. For example a study of diabetic treatment on blood glucose profile over x years may be interested in both the sustained effect of treatment as well as the trajectories of treatment effect over time. Time-varying confounding, where, say, health related measures impact both treatment and outcome after the first recorded time period are a major challenge. Since time-varying confounders often serve as mediators of the effect of previous treatment, simply controlling for them all at once blocks indirect effects of the treatment. To handle this, sophisticated methods such as the g-methods \cite{Hernan} are needed. Such methods however require that there is no unmeasured time-varying confounding.\par
A popular approach to deal with unmeasured confounding is to use Instrumental Variables (IVs), sources of exogenous variation which 1: strongly directly affect treatment assignment, and 2: affect the outcome only through the treatment. IV analysis has had widespread use across multiple disciplines such as genetics, econometrics and clinical studies. Work on IV methods in single IV, time fixed settings is numerous, with the Wald estimator, Two Stage Least Squares (2SLS) and Inverse Variance Weighting among the most notable \cite{Lastref}.\par
This has motivated research in recent years to develop methods that apply IVs to time-varying settings. This is a uniquely difficult challenge, as to identify multiple distinct causal effects, one requires at least as many instruments, when identifying even one instrument can be difficult \cite{Tian2024}. When there are fewer IVs, strict assumptions must be applied to the causal estimand \cite{Joy2022}. One solution is to use a time-varying measure as an instrument.\par
The main challenge with applying IVs in time-varying settings is to identify an instrument that remains strongly associated with treatment assignment over time. Time varying IVs have recently been used in the fitting of Marginal Structural Models using a novel inverse weighting procedure \cite{Cui,Haben}, which used travel distance from the nearest treatment facility as an instrument at two recorded times. Other recent works \cite{Chen2021} used the same time varying instrument to improve Dynamic Treatment Regimes (DTRs) via IVs. One limitation of these methods was relatively simple data structures with regards to the dependence of IV on treatment and confounding. The weighting and DTR approaches only considered a binary IV, dichotomised from a continuous measure, which tends to reduce the strength of the IV. A recent work by the authors \cite{Tompsett} investigated using time varying instruments in a g-estimation setting, based on an adaptation of recent methods \cite{Joy2022}. It showed that a g-estimation approach may perform better than a weighting approach using binary IVs.\par
The papers primary aim is to develop and investigate the application of time varying IVs using Two Stage Least Squares (2SLS). Whilst this has been considered \cite{Robins2000b}, there is, at least to our knowledge, relatively little guidance as to how to apply a time-varying instrument in 2SLS settings. \par
Of the well established IV methods, 2SLS is a strong choice. It's easily used and widely understood in a variety of disciplines, including genomics, biostatistics and econometrics \cite{Angrist,Gudemann,Sanderson2020}. It is also a fairly modular, comprising of two steps that can be adapted and expanded upon to meet several challenges. Crucially, it is the basis of most work in Mendelian Randomisation (MR) studies, of which has seen significant development of 2SLS, using genetic traits as a rich source of multiple instruments at baseline \cite{SBurgess,Misp,Tian2024}.\par
We adapt existing multivariate 2SLS methods in MR \cite{Sanderson2020,Sanderson2,George} and Robust 2SLS methods in time fixed settings \cite{Didelez,Diaz}, to settings with time-varying confounding and evaluate its statistical properties in a simulation study. We compare the relative performance of this approach with standard 2SLS methods and a g-estimation approach. Methods are illustrated in a case study evaluating second line treatment for type 2 diabetes, using GP prescribing preferences for a new class of inhibitors as the time-varying IV. Preference is widely applied as an IV in retrospective clinical studies \cite{Gudemann}, and is naturally time varying in nature. Whilst GP preference was considered previously \cite{Tompsett}, the paper focused on comparison of methods, and the case study had limited scope. We highlight the challenges of operationalising prescription preference as a time-varying IV.

\section{Motivating Example: Comparative Study of Second Line Therapy for Type 2 Diabetes}
Our paper is motivated by an analysis of the effectiveness of second line therapy for Type 2 Diabetes (T2D) on blood glucose levels. T2D is a progressive disease characterized by a impaired ability for pancreatic ${\beta}$-cells to release insulin, leading to elevated glycated haemoglobin (HbA1c), or blood glucose levels. More than 4.4 million people are estimated to be living with T2D in the UK with more than 3.2 million at an increased risk in the future \cite{DUK}, and contributes to increased risks of cardiovascular disease, chronic kidney disease and vascular difficulties.\par
Treatment involves prescribed medication to control and lower HbA1c. In first instance,  NICE (2022) prescription guidelines in the UK recommend Metformin monotherapy. However as patients grow resistant to treatment, a second line intensification is often necessary. Second line therapy supplements Metformin with a second oral anti-diabetic, the choice of which, while guided, is left to clinicians and primary care practices. For this reason, second line therapy preferences can differ greatly between practices and GPs, and is subject to change over time \cite{Curtis}. Of these, patients without high risk of cardio-vascular disease are most commonly assigned Sulfonylureas (SU), or DPP4-inhibitors (DPP4). As one of the first intensifications for Diabetes, GPs tend to have a strong historical preference for SU. However recent studies advocating the advantages of DPP4 may have shifted preference towards DPP4 in recent years \cite{DPP4B,DPP4B2}.

\subsubsection*{Study Population and Eligibility Criteria}
The data comprises of routinely collected primary care data from three East London clinical commissioning groups (CCGs) based in Tower Hamlets, Newham and City, and Hackney. The data was provided by the Queen Mary University of London, Clinical Effectiveness Group (CEG). Data on treatment and health related information was collected and recorded in 6 monthly intervals from 2012 to 2018. Our initial data subset includes $n=7342$ patients who were recorded as initiating second line therapy for diabetes for the first time, after monotherapy failed, between October 2012 to October 2017. This period coincided with a shift in the view of prescribing DPP4 versus SU \cite{DPP4B}.   \par
The median time patients stay in the study after starting second line therapy is 5 follow up periods, or around 2.5-3 years. We look to follow up patients for up to 2 years, or 4 periods. Time $t=1$ is taken as the first period the patient was recorded taking second line therapy, indicating they had begun this treatment within the last 6 months. Times $t=2$ and $t=3$ represent a 6 and 12 months follow up time respectively, with the outcome recorded 6 months later, 18-24 months after initiation.\par
Eligible patients were between 18 and 89 years of age, registered with a primary care practice, and initiated second line therapy after first line monotherapy failed. Patients were required to be on either SU or DPP4 at initiation of second line therapy, with complete relevant data available for the full follow up period. Patients who do not start on one of these two treatment regimes, leaves the study before 3 follow up times, pauses treatment on SU or DPP4, or begins a further intensification by taking both or another diabetic treatment, are censored from the study. There are $n=3640$ patients who meet this criteria with complete treatment data, and $n=2561$ with complete data on all relevant variables.
\subsubsection*{Treatment Comparison}
Treatment is a contrast of one of two first intensification second line treatments.\par
Treatment: Initiate treatment intensification with DPP4 over 18 months.\par
Control:  Initiate treatment intensification with SU over 18 months. \par
Treatment is recorded at each 6 month interval, with the "Treatment" group denoted 1, and the "Control" group, denoted 0.
\subsubsection*{Outcome}
Outcome: Outcome is the recorded measure of HbA1c levels in mmol/mol 18-24 months after initiation of second line therapy. 
\subsubsection*{Objectives}
The estimand of interest  is the Average Treatment Effect (ATE) of sustained treatment with DPP4, compared to SU, over 18 months. Of secondary interest is the how the the effect of treatment changes over 6 monthly intervals. Due to the risk of unmeasured confounding, we use a measure of physicians prescription preference (PP) taken over time as an IV. Full details are in the methods section.
\subsubsection*{Covariates}
Information on covariates measured at baseline is summarised in Table 3 in the Appendix. Data is available on age, gender, ethnicity, HbA1c levels, Body Mass Index, systolic blood pressure, blood lipid profiles, kidney function, and history of stroke and hypoglycemic events. Co-prescription history of statins and beta blockers was also available. No patients were prescribed Insulin in this study. We consider that prescription preference may depend on some of these confounders. Patients on DPP4 have lower HbA1c levels at baseline and prior to 2nd line therapy, with higher levels of Body Mass Index over 34. Notably, patients were majority non-white, with around 75\% recorded as Black, South Asian, or Other Ethnicity.

\section{Methods}
\subsection{Overview}
Suppose $T$ time periods for which we observe a treatment $A_t$ up to time $T$. For the rest of the paper we will refer to "treated" $(A_t=1)$ and "control" $(A_t=0)$ groups. When referring to variables with no subscript $t$, this is defined as the set of all observations of that variable, that is $A=(A_1,\ldots,A_t)$ etc. We observe a continuous end of study outcome $Y$ at time $T+1$, and observed, and unobserved time varying confounders $L_t$ and $U_t$, confounding the effect of $A_t$ on $Y$. We define $Z_t$ as the time-varying instrument.\par
We represent our expected data setups in the Directed Acyclic Graph (DAG) shown in Figure 1. Previous works with time dependent variables \cite{Chen2021,Haben} assumed similar data structures with the red directional arrows removed. We refer to this as the "Simple" data setup. A more "Complex" setup includes these arrows, and is akin to structures discussed in \cite{Cui,Tompsett}.\par
Uniquely difficult aspects of this setup is the introduction of time-dependent $Z-A$ confounding, direct associations between instrument times, and a much more complicated $Z-A$ relationship, whereby $A_t$ is both directly influenced by $Z_{t-1}$, and vice versa. These relationships may create problems for standard time varying causal methods as we explain below.\par
Taking our motivating example, $Z_t$ is preference for prescribing DPP4 over SU over the period $t$ (PP). We expect PP to be correlated over time. We might also expect that certain time varying health related measures $L_t$, progression of disease, blood glucose levels or patient self reported satisfaction, could drive a change in which drug is preferred over time beyond new emerging research and word of mouth. Moreover, PP is likely to be dictated by prescriptions given over the long term, particularly for patients with a long history of treatment. Arrows running from past PP to future treatment may occur when repeat prescriptions are handed out, and as such preference at initiation remains a direct factor on assignment. We therefore expect that two stage methods to handle the complex setup of the DAG are needed.

\begin{figure}[H]
\centering
\begin{tikzpicture}[>=stealth,semithick]
\node[] (l1) at (0,  12)   {$L_1$};
\node[] (l2) at (4,  12)   {$L_2$};
\node[] (l3) at (8,  12)   {$L_3$};

\node[] (z1) at (2.0,  8.5)   {$Z_1$};
\node[] (a1) at (2.0,  10.5)   {$A_1$};
\node[] (z2) at (6.0,  8.5)   {$Z_2$};
\node[] (a2) at (6.0,  10.5)   {$A_2$};

\node[] (z3) at (10.0,  8.5)   {$Z_3$};
\node[] (a3) at (10.0,  10.5)   {$A_3$};

\node[] (y4) at (12,  12)   {$Y$};

\node[] (u1) at (0,  13)   {$U_1$};
\node[] (u2) at (4,  13)   {$U_2$};
\node[] (u3) at (8,  13)   {$U_3$};

\path[->] (z1) edge node {} (a1);

\path[->,red] (z1) edge node {} (a2);

\path[->,red] (a1) edge node {} (z2);

\path[->] (l1) edge node {} (l2);

\path[->] (l2) edge node {} (l3);

\path[->] (z2) edge node {} (a2);

\path[->,red] (a2) edge node {} (z3);

\path[->,red] (z2) edge node {} (a3);

\path[->] (z3) edge node {} (a3);

\path[->] (z1) edge node {} (z2);

\path[->] (z2) edge node {} (z3);

\path[->] (a1) edge node {} (a2);

\path[->] (a2) edge node {} (a3);

\path[->] (l1) edge node {} (a1);

\path[->,dotted] (u1) edge node {} (a1);
\path[->,dotted] (u2) edge node {} (a2);
\path[->,dotted] (u3) edge node {} (a3);

\path[->,dotted,red] (l1) edge node {} (z1);
\path[->,dotted,red] (l2) edge node {} (z2);
\path[->,dotted,red] (l3) edge node {} (z3);

\path[->] (a1) edge node {} (l2);

\path[->] (l2) edge node {} (a2);

\path[->] (a2) edge node {} (l3);

\path[->] (l3) edge node {} (a3);

\path[->,dotted] (a1) edge node {} (y4);
\path[->,dotted] (a2) edge node {} (y4);
\path[->] (a3) edge node {} (y4);

\path[->,dotted] (l1) edge [bend left = 10] node {} (y4);
\path[->,dotted] (l2) edge [bend left = 5] node {} (y4);
\path[->] (l3) edge node {} (y4);

\path[->]  (u1) edge node {} (l1);
\path[->]  (u2) edge node {} (l2);
\path[->]  (u3) edge node {} (l3);

\path[->]  (u1) edge [bend left = 30] node {} (y4);
\path[->]  (u2) edge [bend left = 25] node {} (y4);
\path[->]  (u3) edge node {} (y4);

\end{tikzpicture}
\caption{DAG of Complex data setup with $T=3$ treatment periods}
\end{figure}
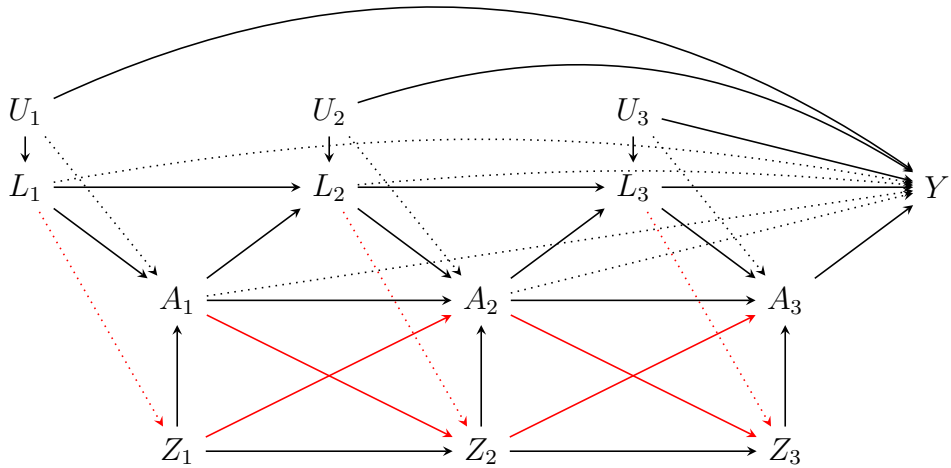
Define $Y(a)$ as the counterfactual outcome that would have been observed under some treatment regime $a=(a_1,\ldots,a_T)$. A causal effect is sought by estimating the contrast in counterfactual outcomes under different treatment regimes. We define this contrast as the Average Treatment Effect (ATE), the average difference in counterfactual outcome in the population when always treated, versus never treated
$$E[Y(1,1,1)-Y(0,0,0)].$$
As with all causal methods, we make the assumptions of counterfactual consistency and positivity throughout the paper \cite{Joy2022}.\par
In time fixed settings, a common alternative contrast is the Local Average Treatment Effect (LATE). This is defined as the effect of treatment in those for whom a change in IV status induces a change in treatment status. When $Z$ is defined as assignment to treatment, and $A$ treatment taken, this is interpreted as the ATE in compliers, those who always take the treatment they are assigned to. We do not consider the LATE in this paper, as it requires certain monotonicity conditions on the data. These conditions, as well as what exactly constitutes a complier, are not well defined, if at all, with a time-varying treatment \cite{Mogstad}.
\subsection{IV Assumptions}
In time-varying settings we make the following specific assumptions for the $Z_t$ to be valid instruments, and are multivariate extensions of those in time fixed settings. Define $M_t$ as the set of variables for which, when conditioned upon, $Z_t$  satisfies the following assumptions.
\begin{enumerate}
\item IV1 (IV Relevance): There exists a measurement of $Z$ and $A$ at each time point, so that we have as many instruments as treatment times, and there exists an association between $Z_t$ and  $A_t$, conditional on $M_t$, at each time point. This association must sufficiently strong.
\item IV2 (Exclusion Restriction): $Z_t \ci Y| M_t\; \forall t$. 
\item IV3 (Conditional Exchangeability): Provided there are no unmeasured confounders between $Y$ and $Z$, we then have that $Z_t \ci U | M_t\; \forall t $, that is $Z$ independent of any other unmeasured confounding bewteen $A$ and $Y$. IV2 and IV3 can be formalised as the assumption
$Y(a) \ci Z_t | M_t\; \forall \; t \; and\; a$.
\end{enumerate}
For our DAG we have in the simple case $M_t=Z_{t-1}$, and for the complex case $M_t=(A_{t-1},Z_{t-1},L_t)$. We define $M$, as the union of these sets for all $t$.
\subsection{Subtantive Models}
Under these assumptions, established works in 2SLS settings present the substantive model of interest as a Linear Structural Mean Model \cite{Didelez}, which models the observed outcome Y as
\begin{equation}
E(Y|A,Z,M)=E[Y_0|A,Z,M]+\sum_{t=1}^T\beta_t A_t
\end{equation}
where $Y_0=Y(0,\ldots,0)$, and our interest lies in the effects of $A_t$ on outcome $\beta_t$. We cannot estimate the ATE under this model without additional assumptions. Firstly, we make the assumption in this paper that the treatments $A_t$ affect $Y$ linearly, with no interaction terms between treatment times. Secondly, we require that the shift in conditional mean outcome is the same in the treated arm, as in the control arm
$$E[Y(\bar{a}_t,0)-Y(\bar{a}_{t-1},0)|\bar{A}_{t-1}=\bar{a}_{t-1}, A_t=1,\bar{Z}_t, M_t]=E[Y(\bar{a}_t,0)-Y(\bar{a}_{t-1},0)|\bar{A}_{t-1}=\bar{a}_{t-1}, A_t=0,\bar{Z}_t, M_t].$$ 
for all $t$, where $\bar{A}_{t}$ represents the history of treatment up to time t. This condition was described in \cite{Robins2006} and more recently in \cite{Tompsett}. It can be shown that this property holds under the "no current treatment interaction property", that is there is no effect modification of the effect of $A_t$ on $Y$ by other variables, a property that clearly holds when the casual effect is linear in $A$. When this property holds, the ATE can be estimated as the sum of the coefficients $\beta_t$. 

\subsection{Two Stage Least Squares Methods}
\subsection*{Standard 2SLS}
The standard 2SLS methodology that is well known in time fixed situations \cite{Hernan} was generalised to the case of multiple exposures periods (or multiple separate exposures) via Multivariate Mendelian Randomisation (MVMR) \cite{Sanderson2,Joy2022}. This applied 2SLS to the case of a time varying exposure, with at least as many measured IVs at baseline. In this paper we have a just-identified situation, where we have as many instruments as time periods. Standard 2SLS can be performed by the following steps.  
\begin{enumerate}
\item First Stage Models: For each $t$, postulate and fit a series of first stage models for each $A_t$ $f_{A_t}=P(A_t|Z,L^{M})$ by fitting a main effects Ordinary Least Squares (OLS) model for  $A_t$ on all instruments $Z_t$ and confounders for which $Z_t$ depend $L^{M}$.
\item Obtain predicted values for each $A_t$ from the fitted models, defined as $\hat{A_t}=E(A_t|Z_1,\ldots,Z_t,L^{M},\alpha^*_t)$, where $\alpha^*_t$ are the fitted coefficients.
\item Second Stage Model: Postulate and fit a main effects OLS model for $Y$ against the predicted values of $\hat{A}_t$ and $L^{M}$.
$$f_{Y}=E(Y|\hat{A}_1,\ldots,\hat{A}_t;\beta)=\beta_{0}+\sum_{i=1}^{T}\beta_{t} \hat{A}_t+\beta_L L^{M}.$$ 
using OLS regression. The estimated coefficients $\beta_t^*$ of the fitted model are the causal effects of interest of Equation (1), for which the ATE can be derived.
\end{enumerate}
With time-varying IVs we expect that standard 2SLS can attain unbiased estimates of causal effects in the simple data setup of Figure 1, without the red arrows. In this instance, the $Z_t$, can be thought of functionally as correlated IVs at baseline, which was demonstrated to perform well in recent work \cite{Joy2022}. We anticipate however that standard 2SLS may be unsuited to the more complex data setup.\par
Firstly, we suspect that 2SLS may not be able to address time-varying confounding between the $Z_t$, $A_t$ and $Y$. In cases of confounding only at baseline, 2SLS can control for this by including any baseline confounders in the first and second stage models \cite{Didelez}. With time-varying confounding, conditioning on confounders beyond baseline blocks indirect causal effects of $A$ on $Y$ through $L_t$, as well as risking collider bias. As such, the third IV assumption, exchangeability cannot be satisfied as a uncontrolled for pathway exists from $Z_t$ to $Y$ via $L_t$.\par
A second problem lies with the associations between $A$ and $Z$, as they induce collider bias into the first stage models. In Figure 1 for example, once the first stage model for $A_1$ conditions on $Z_2$, it collides, and subsequently bias, the $Z_1-A_1$ association. 

\subsection*{Robust 2SLS (R2SLS) For Time Varying Instruments}
To remedy this problem we now present the main method of the paper. We turn to Robust 2SLS methodologies \cite{Okui,Didelez,Diaz}. These methods deal with instances in time fixed scenarios where exhangeability of $Z$ is dependent on some unmeasured confounder at baseline $V$. These methods propose that before the first and second stage model is fitted, a parametric model for the instrument $Z$ is postulated
$$f_{Z}=E(Z|V).$$
A estimating equation method is then described that equates to the following procedure.
\begin{itemize}
\item Fit model $f_{Z}$, and calculate predictions of $Z$ as $Z^*$
\item Attain the residuals of $f_{Z}$, $Z-Z^*$
\item Perform standard 2SLS, with $Z-Z^*$ in place of $Z$, controlling for $V$ in the first and second stage models.
\end{itemize}
To handle the complex data structure of Figure 1, we consider the following extension of these methods to a time varying setting.
\begin{enumerate}
\item Postulate a model for each $Z_t$ given $M_t$
$$f_{Z_t}=E(Z_t|M_t;\gamma)$$
using an appropriate regression model. Fit this model and calculate predictions for $Z_t$ as
$$\hat{Z}_t=E(Z_t|M_t;\gamma^*)$$.
\item From this define residuals $Z_t^{res}=Z_t-\hat{Z}_t$
\item Now perform 2SLS, replacing $Z_t$ with $Z_t^{res}$. 
\item Standard 2SLS may be perform  without conditioning on $L_t$ in the first and second stage models.
\end{enumerate} 
One way to think of this method is that it is an application of standard 2SLS, taking $Z_{t}^{res}$ as the time-varying IVs of the method. These residuals instruments are sometimes known as the Frisch-Waugh-Lovell residualised instruments, that are designed to allow the fitting of partial regression models, and are orthogonal (unassociated) with $M_t$. This property is crucial for two reasons. Firstly it "models out" the associations between $A_{t-1}$ and $Z_{t}$, and thus removes the collider bias they induce in the first stage model. \par 
Secondly we have that $Z^{res}_t | L_t$ and thus there is a choice as to whether to control for $L^{M}$ in the first and second stage models at all.
In the complex case this means we no longer need to control for time-varying confounding in the first and second stage models which blocked indirect effects of $A_t$ on $Z_t$. We can thus obtain unbiased estimates of the causal effects in the complex case provided that $f_{Z_t}$ is correctly specified.\par
These methods have been referred to as g-estimators \cite{Diaz}, and in fact R2SLS is equivalent to to the g-estimation procedure recently considered in \cite{Joy2022}, and \cite{Tompsett} for time varying instruments, when the first and second stage are fit using OLS models. We show this in the appendix.
\subsection*{Robustness properties of R2SLS}
We take specific note of the use of R2SLS in the simpler data setup. Recent work \cite{Didelez} stress tested this methodology in the single time period case, in situations in which $A_t$ and $Y$ depend on non linear relationships, such as squared terms of $L_t$ and noted that the method was doubly robust. Provided that either $f_{Z_t}$ or $f_{Y}$ was correctly specified.\par
This robustness however does not extend to the complex case, where we use R2SLS as we cannot control for confounding beyond baseline in $f_{Y}$. In essence we are leveraging the doubly robust property to avoid modeling problematic associations in the first and second stage models. But in doing so, we are reliant on correct specification of $f_{Z_t}$. How vulnerable this method is to misspecification of $f_{Z_t}$ is a point of interest.\par
One consideration is what type of linear models should be used. We consider situations in with $f_{Z_t}$ is fit by its natural linear model and the first and second stage are fit by Ordinary Least Squares (OLS). When $A_t$ is binary, this corresponds to a Linear Probability Model (LPM). The appendix considers other popular choices of first stage model in the case of binary treatment.

\section{Simulations}
\subsection{Data generating mechanism}
We will test standard 2SLS and R2SLS via simulation. The major objectives of the simulation are firstly, to determine the consistency and efficiency of standard 2SLS, R2SLS-L and R2SLS under the data setups of Figure 1 under varying IV strengths and sample sizes. Secondly, we aim to test the robustness of R2SLS in simple and complex situations in which the $Z_t$, $A_t$ and $Y$ depend on square terms of confounders, or $Z-A$ interaction terms, leading to misspecification of $f_{Z_t}$. The data is simulated as follows.
\begin{itemize}
\item The time-varying instrument $Z_t$ is simulated as $logit(P(Z_t=1))=\mu_{Z_t}-E(\mu_{Z_t})$ where in the simple case
$$\mu_{Z_t}=Z_{t-1}$$
and the complex case as
$$\mu_{Z_t}=Z_{t-1}+A_{t-1}+3*L_{t}+\sigma_{Z}(L_{t}^2).$$
Here $\sigma_{Z}$ introduces typical terms that may trigger misspecification of $f_{Z_t}$. 
\item $A_t$ is a binary variable, generated using $$logit(P(A_t=1))=\Phi(\mu_{A_t}-E(\mu_{A_t}))(1-\Delta_t)+Z_t\Delta_t.$$
where
$$\mu_{A_t}=A_{t-1}+L_{t}+U_{t}$$
in the simple case and 
$$\mu_{A_t}=Z_{t-1}+A_{t-1}+L_{t}+U_{t}+\sigma_{A}(L_{t}^2)$$
in the complex case with $\Phi$ denoting the standard normal cdf. We use $\sigma_{A}$ to introduce non linear terms and interactions into the models. We simulate this way based on work in \cite{Haben,Tompsett} where $\Delta_t$ can be chosen to control the strength of the instrument $Z_t$, simulated as $\Delta_{t}=\Phi(\alpha)$.
\item Lastly $Y$ is normally distributed with variance 1, and mean
$$E(Y)=\sum_{i=1}^T (U_t+A_t+L_t)+\sigma_{Y}L_1^2$$
\end{itemize}
The true values for $\beta_t$ are $(3,2,1)$, and subsequently the ATEs true value is 6.
\subsection{Implementation}
In each simulation we generate 1000 datasets. Confidence intervals (CIs) are obtained via a percentile bootstrap method using 1000 bootstrapped datasets. In all simulations standard 2SLS is performed as in section 3, controlling for $L_1$. R2SLS is performed with $f_{Z_t}$ defined as main effects logistic regression model on $Z_{t-1}$ for the simple case and, in the more complex case $Z_{t-1}$, $A_{t-1}$ and $L_t$. $f_{A_t}$. The first stage model is a main effects LPM regression of $A_t$ on $Z^{res}$, and the second stage model is a main effects OLS regression of $Y$ on $\hat{A}$ \par
We look to vary the sample size $n$ which will be set to 1000 or 5000, and the parameter $\alpha$, which can be set to values between 0 and 1 to influence the strength of association between $Z_t$ and $A_t$. We choose these values as $0.1$, $0.3$ and $0.5$, which correspond to weak, moderate, and strong associations respectively.\par
We also look to test the robustness of R2SLS in the complex case, which cannot use doubly robust methods to misspecified models, when we set $\sigma_{Z}$ and $\sigma_{Y}$ to $0$ or $1$ respectively. When non-zero, these terms introduce non linear terms into $f_{Z}$ and $f_{Y}$, specifically square terms of the baseline confounder $L_1$ and interactions between past treatment and instrument, leading to misspecified models.\par
As recommended in \cite{MorrisMCE} we report for the ATE the average absolute bias, Root Mean Square Error (RMSE) and Coverage defined as the proportion of datasets in which the bootstrap CI included the true value.

\subsection{Results}
Tables 1 and 2 present the results of the simulations in the case of no non-linear terms $(\sigma_Z,\sigma_A,\sigma_Y)=(0,0,0)$. The full simulation results are in the Appendix, Tables 4, 5 and 6. We can see that standard 2SLS shows the same effectiveness as the robust case in the setup for our simpler simulation. This is expected, and identified by the work in \cite{Joy2022}. With the simple setup of Figure 1, the $Z_t$ can be treated as though they were baseline instruments. Results show low bias and good coverage in all cases where instrument strength is acceptably strong. In table 4 2SLS was employed using a first stage probit model rather that OLS, which also performed well. Results began to show bias only at low correlation strengths $\alpha=0.1$ and low sample size $n=1000$. \par
As expected in complex scenarios with time varying confounding R2SLS is the only method capable of obtaining unbiased estimates conditional on sufficient instrument strength. 2SLS shows high bias in all cases. Probit based first stage models in Table 5 no longer performed well, even for R2SLS. This may infer that the orthogonality to $M_t$, for which R2SLS is based upon is not guaranteed when using a non OLS model. It also suggests that unaccounted for time varying confounding can lead to misleading results. We can infer from Tables 1 that R2SLS with LPMs is the only reliable method available in complex circumstances.\par
Table 2 shows the results using R2SLS for the complex data setup when we add non linear terms to the models. We add terms into the models for $Z$ and $A$ at the same time, in order to control the instrument strength. As expected, when all 3 models are misspecified we encounter biases and less efficient results in all cases. This bias is quite low however in cases of strong instrument strength and sample size.\par
When $\sigma_{Z}=1$ and $\sigma_{Y}=0$ results show bias similar to when both are 1. This is as expected as the inability to model the Y-M association means there is not robust protection from misspecified $f_{Z_t}$. The opposite way around however results are unbiased, subject to instrument strength, as correct specification of $f_{Z_t}$ protects from misspecification of $f_{Y}$, as expected.

\begin{table}[H]
\begin{center}
\begin{tabular}{|c|c|c|c|c|c|}
\hline
 n &  $\alpha$ & Bias & RMSE & MCE &Coverage\\
\hline
\multicolumn{6}{|c|}{Simple Simulation}\\
\hline
\multicolumn{6}{|c|}{Standard 2SLS}\\
\hline
5000 & 0.5 & 0.004 & 0.42& 0.006 & 93.5\\
5000 & 0.3 & 0.012 & 0.52& 0.009 & 93.6\\
5000 & 0.1 & 0.096 & 0.85& 0.022 & 94.2\\
1000 & 0.5 & 0.013 & 0.63& 0.012 & 94.5\\
1000 & 0.3 & 0.048 & 0.79& 0.020 & 95.6\\
1000 & 0.1 & 0.128 & 4.18& 0.554 & 98.9\\
\hline
\multicolumn{6}{|c|}{R2SLS (g-estimation)}\\
\hline
5000 & 0.5 & 0.004 & 0.42& 0.006 & 93.5\\
5000 & 0.3 & 0.012 & 0.52& 0.009 & 93.6\\
5000 & 0.1 & 0.096 & 0.85& 0.022 & 94.2\\
1000 & 0.5 & 0.013 & 0.63& 0.012 & 94.5\\
1000 & 0.3 & 0.048 & 0.79& 0.020 & 95.6\\
1000 & 0.1 & 0.128 & 4.18& 0.554 & 98.9\\
\hline
\multicolumn{6}{|c|}{Complex Simulation}\\
\hline
\multicolumn{6}{|c|}{Standard 2SLS}\\
\hline
5000 & 0.5 & 1.900 & 1.38& 0.003 & 0.0\\
5000 & 0.3 & 2.513 & 2.81& 0.238 & 76.3\\
5000 & 0.1 & 3.397 & 1.91& 0.041 & 2.7\\
1000 & 0.5 & 1.895 & 1.38& 0.007 & 0.2\\
1000 & 0.3 & 2.781 & 4.12& 0.531 & 78.3\\
1000 & 0.1 & 3.361 & 2.87& 0.240 & 40.9\\
\hline
\multicolumn{6}{|c|}{R2SLS (g-estimation)}\\
\hline
5000 & 0.5 & 0.010 & 0.33& 0.003 & 96.1\\
5000 & 0.3 & 0.014 & 0.42& 0.006 & 95.2\\
5000 & 0.1 & 0.638 & 3.21& 0.325 & 96.2\\
1000 & 0.5 & 0.006 & 0.50& 0.008& 95.8\\
1000 & 0.3 & 0.023 & 0.644& 0.013 & 95.4\\
1000 & 0.1 & 5.352 & 11.439& 4.140 & 98.7\\
\hline
\end{tabular}
\caption{Simulation results, targeting the ATE with coverage based on b=1000 bootstrapped samples.}
\end{center}
\end{table}

\begin{table}[H]
\begin{center}
\begin{tabular}{|c|c|c|c|c|c|c|c|}
\hline
$\sigma_{Z}$, $\sigma_{A}$ & $\sigma_{Y}$  &n &  $\alpha$ & Bias & RMSE & MCE &Coverage\\
\hline
\multicolumn{8}{|c|}{R2SLS (g-estimation}\\
\hline
1 & 1 &5000 & 0.5 & 0.026 & 0.60& 0.011 & 94.2\\
1 & 1 &5000 & 0.3 & 0.083 & 0.75& 0.017 & 93.7\\
1 & 1 &5000 & 0.1 & 0.690 & 2.22& 0.154 & 98.3\\
1 & 1 &1000 & 0.5 & 0.101 & 0.90& 0.026& 95.5\\
1 & 1 &1000 & 0.3 & 0.192 & 1.17& 0.042 & 94.6\\
1 & 1 &1000 & 0.1 & 6.922 & 12.41& 4.868 & 98.3\\
\hline
1 & 0 &5000 & 0.5 & 0.013 & 0.63& 0.012 & 93.7\\
1 & 0 &5000 & 0.3 & 0.016 & 0.77& 0.018 & 94.6\\
1 & 0 &5000 & 0.1 & 0.593 & 2.36& 0.176 & 97.6\\
1 & 0 &1000 & 0.5 & 0.075 & 0.95& 0.028& 94.0\\
1 & 0 &1000 & 0.3 & 0.165 & 1.22& 0.047 & 94.6\\
1 & 0 &1000 & 0.1 & 7.000 & 12.42& 4.877 & 98.2\\
\hline
0 & 1 &5000 & 0.5 & 0.010 & 0.33& 0.003 & 95.7\\
0 & 1 &5000 & 0.3 & 0.015 & 0.42& 0.006 & 95.1\\
0 & 1 &5000 & 0.1 & 0.228 & 1.45& 0.066 & 96.5\\
0 & 1 &1000 & 0.5 & 0.002 & 0.50& 0.008& 95.7\\
0 & 1 &1000 & 0.3 & 0.01 & 0.64& 0.012 & 95.5\\
0 & 1 &1000 & 0.1 & 3.440 & 12.20 & 4.709 & 98.8\\
\hline
\end{tabular}
\caption{Simulation results targeting the ATE, using R2SLS (g-estimation) for complex data setups with mispecified models for $Z$ $A$ and $Y$.}
\end{center}
\end{table}

\section{Case Study}
\subsection{Instrument Definition: GP Prescription Preference}
Our instrument is a measure of GP prescription preference (PP) for DPP4 over SU over time. There are 139 GPS in the data, with an average of 27 patients each, ranging from 1 to 98 patients. A recent paper \cite{Gudemann} summarized well the various measures to approximate GP preference via proportion of prescriptions issued. Available prescription data does not include specific dates, so subject specific measures of PP are not calculable. We instead consider GP specific measures of preference at each 6 month period based on the definitions of $IV_{prop}$ \cite{Gudemann}. 
\begin{enumerate}
\item $PP^{Cal}$: A measure of GP preference at each 6 monthly calendar period is taken as the proportion of all prescriptions of DPP4 as second line treatment within that 6 monthly calendar period. An individuals value of $PP^{Cal}$ at some follow up time $t$ is then GP preference during the 6 months calendar period when $t$ occured. We label this $PP_{t}^{Cal}$
\item $PP^{t}$: Alternatively, a measure of PP over follow up time $t$, rather than calendar time can be considered. The proportion of a GPs prescriptions at initiation are taken as $PP_{1}^{t}$. This is repeated for follow up times 2 and 3.  This represents a measure of how GP may change preference based on how long a patient has been taking second line treatment.
\end{enumerate}
A visual representation of the definitions of $PP$ is shown found in Figure 3 in the Appendix. Preference taken at 6 month period shows a clear shift in preference over time towards DPP4, with a proportion of around 15\% in 2013 to 40\% in 2017 . However, $PP^{t}$ shows minimal change between follow up periods.
\subsubsection*{IV1: IV Relevance}
It is important in a practical context to investigate if $Z_t$ satisfy the three main IV assumptions. Unfortunately, only IV1, IV relevance can be tested from the available data. Few patients switch from one treatment to another over the study period. We need to investigate if there is firstly sufficient strength of association between PP and treatment at initiation, and then assess if there is strong enough change in this association over time for multiple instruments. \par
In a time fixed setting, the Cragg-Donaldson F-statistic is used to evaluate the strength of instruments \cite{Cragg,Yogo1,Yogo2}. An F-statistic of 10 is historically cited as a sign of a sufficiently strong instrument, however more recent work suggesting a value of 100 may be necessary \cite{Moler}. This test is not however appropriate to evaluate the overall strength of multiple IVs. The Sanderson-Windmaijer conditional F-test \cite{Sanderson2,FTEST}, adapts the Cragg-Donaldson statistic to provide a value for each $A_j$, testing the remaining predictive power in the $Z_j$ once all other $A_j^{res}$ have been calculated in the first stage model. Analysis of standard errors, unusual estimates, and inspecting the influence of $Z_t$ in the first stage models, can also infer whether there is sufficient instrument strength.  \par
The first stage F-statistics for PP at $t=1$ were 1422 and 927 for $PP^{Cal}$ and $PP^{t}$ respectively. Strengths similar to this were found in a recent paper on PP in assessment of T2D second line treatment \cite{Bidulka}. The conditional F-tests for $PP^{Cal}$ and $PP^{t}$ however ranged between 30 and 70, below the recently recommended standard of 100, and dramatically lower then the first stage F-statistics.  Correlations between PP times were very high, above 0.85 for $PP^{Cal}$, and in the case of $PP^t$ they neared 1. This explains the heavy drop in predictive power and indicates a high risk of collinearity.
\subsubsection*{IV2: Exclusion Restriction}
It cannot be determined from the available data if there is a direct association between GP preference and HbA1c levels at any future time. One possible avenue is that GPs with a preference for SU over DPP4, may be of "higher standard", leading to an improvement in HbA1c levels. Whilst this assumption cannot be tested, it can be reasoned that standard of care between practices will depend on numerous other more critical factors such a funding, region and standards of follow up care, and such a possibility is unlikely.
\subsubsection*{IV:3 Exchangeability}
GP preference may be affected by past confounders, and measurements of outcome. Dependence on past preference and treatment history is likely, but can be easily controlled. However a GP identifying poor past performance of patients on one drug may switch preference as a result. This opens the possibility of preference drifting over time towards DPP4 being dependent on other health related measures other than past treatment, such as HbA1c history. We do not for example, have full information on comorbidities. However, the preference of a GP may be reinforced by the overall health of their previous patients prior to that 6 month period. We argue that health related issues are normally considered on a patient by patient basis by GPs, and thus should directly affect treatment assignment, rather directly impacting preference.  It is also unlikely for exchangeability to be met at initiation, only to fail at later follow up times, as health related factors are always being considered. It is most crucial therefore to be confident of exhangeability at the first time period.

\subsection{Estimating Approaches}
We use 2SLS and R2SLS to investigate the effect of sustained treatment with DPP4 versus SU on HbA1c levels at 18 months, estimating the Average Treatment Effect and as a secondary measure, the trajectory of treatment effect over time. We repeat these analyses for $PP^{Cal}$ and $PP^{t}$. The models of interest are as follows.
\begin{itemize}
\item Standard 2SLS: The Standard 2SLS as described in Section 3.4. This represents a benchmark analysis, as we anticipate that it will attain biased causal effects.
\item R2SLS:  The Robust 2SLS methodology described in Section 3.4.
\item  Ridge R2SLS: The R2SLS method in which the second stage model is fit via Ridge regression methods to address collinearity. See Appendix for details
\end{itemize}
Confidence intervals are obtained by non parametric bootstrapped with $b=1000$ resampled datasets.\par
R2SLS models $Z_t$ against history of treatment, and instrument as well as history of HbA1c levels, Median HbA1c levels prior to initiation, smoking status, calendar period and ethnicity. These were identified as potentially unbalanced by PP at $t=1$ (see Appendix, Table 8).\par
Ridge R2SLS applies ridge regularisation penalty for the second stage in an attempt to deal with multicollinearity in the second stage model. An algorithm is described in the appendix, based on ridge trace plot stabilising efforts, to identify a penalty for which the $\beta_j$ were all within 0.01 of the values at the previous tested penalty.

\subsection{Results}
A full table of results are shown in the Appendix, Table 7. We show the results for the ATE in Figure 2. All methods suggest a reduction in HbA1c levels with sustained treatment with DPP4 compared to SU. Standard 2SLS identified a decrease in HbA1c levels at 18 months of around  5 mmol/mol. Using R2SLS, this decrease was between 2.5 to 3 points. All methods indicated a significant reduction in HbA1c levels. \par
There was a modest gain in precision with RS2LS over standard 2SLS, and a reduction in the effect size. This may indicate that 2SLS was affected by collider bias or unmeasured confounders of the IV. Results with $PP^{Cal}$ were also slightly more precise than with $PP^{t}$.\par
Whilst R2SLS was able to identify an stable estimate of the ATE, trajectories however suffered from very unstable estimates and confidence intervals. Inspection of the first stage models indicate that PP at time 1, explains the vast majority of the treatment-PP association over time, leading to jointly weak instruments. The second stage model also displayed very high collinearity, with a Variance Inflation Factor (VIF) in the 100s or 1000s, which will heavily inflate standard errors.
\par
Use of Ridge R2SLS soothes the high instability caused by collinearity and pulls estimates towards the null. The ATEs are similar with a modest reduction in the width of the confidence interval. Trajectories imply a dampening of the effect of DPP4 over time. However trajectories using this method are likely to remain biased due to weakly joint instruments, an issue which Ridge regularizations are unsuited to address. Stabilising estimates with Ridge regularisation requires further work before being recommended more generally. This is considered in the discussion.

\begin{figure}[H]
\begin{center}
\includegraphics[width=1\textwidth]{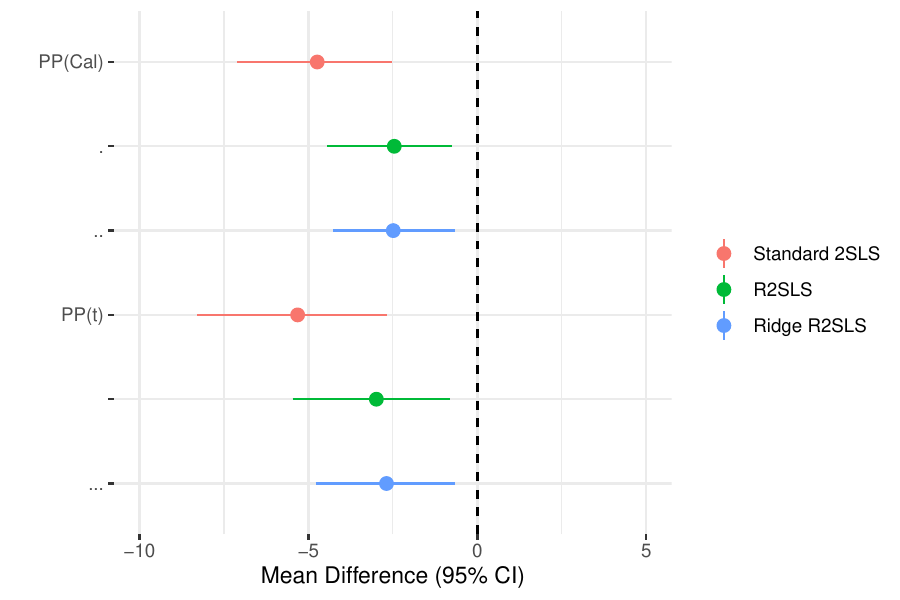}   
\caption{Forest Plot of case study results for PP}
\end{center}
\end{figure}

\section{Discussion}
In this paper, we investigate the use of Standard and Robust 2SLS methodologies in a full time-varying setting. We find by simulation that Standard 2SLS can perform well in situations in which the IV has simple relationships between treatments and confounders, but showed biased estimates in more complicated scenarios. Robust 2SLS was found to perform well even in complicated setups. Results for R2SLS showed bias only at weak correlation strengths and low sample size. R2SLS also performed relatively well in when non linear terms were introduced. An evaluation of the effect of sustained treatment intensification with DPP4 versus SU over 18 months found a significant reduction in HbA1c levels of between 2.5-5 mmol/mol.\par
This paper contributes to the literature in the following ways. We critically evaluate the performance of 2SLS methodology in a full time-varying setting with time-varying confounders, instruments, and complex relationships over time. We adapt literature on robust 2SLS methods in time fixed situations to a multivariate setting and demonstrate that this adaptation can successfully recover time varying causal effects by modeling dependence on prior time periods. We identify that R2SLS can be related to recent advances in g-estimation for time-dependent instruments. We also assess the performance of R2SLS in multivariate setting under non linear relationships between the IV, treatment, and time varying confounders.\par
We build on prior literature focusing on Multivariate Mendelian Randomisation \cite{Sanderson2} by extending its methodologies to time-varying instruments and time varying-confounding. We consider practical situations rarely encountered in genetic studies and find the methods can handle these data setups. We extend the existing work on Robust 2SLS methods detailed for time fixed scenarios \cite{Diaz,Didelez}, and find that they are suitable to handle time varying confounding, and retain the robust properties, identified in the univariate setting. We also confirmed that the relative robustness to non linear relationships appear to extend to multiple time periods, provided the model for the IV is correctly specified.\par
Recent studies have compared SU versus DPP4 in both a clinical trials setting \cite{Head2Head1,Head2Head2}, and a retrospective cohort \cite{Fadini}.
In particular, recent work using IVs compared glycemic control in SU and DPP4 using physician preference as a baseline instrument \cite{Bidulka,Gudemann}. In these studies, around 75\% of patents were white. In this paper, our study was 75\% non-white, specifically patients of South Asian and black ethnicity. Prevalence of diabetes is estimated to be two to four times higher in Asian and Black Ethnic Groups \cite{Lancet}, and there is as such considerable interest in the external validity of previous studies on glycemic control to non-white populations in the UK. We identify effects of DPP4 versus SU that are in line with these previous studies, providing external validity to these papers and evidence of the effectiveness of DPP4 in non white populations.\par
Simulations showed that whilst non linear and interaction terms in model for the instrument are relatively robust to bias there is, as is typical for IV methods, a heavy penalty if the main effect of a confounder of the $Z-Y$ relationship is not controlled for. Unlike time fixed R2SLS the outcome model can rarely be used as a means to make results doubly robust. Heteroskedastic errors in the instrument may also impact the ability of the method to "model out" complex associations between $Z$ and $A$.\par
For the case study, we found that even with identifying PP with strong strength at baseline, it was challenging to attain time-varying PP with enough variation in association over time to identify treatment trajectories. This was due to high correlations between the instruments. Specifically, PP at initiation contained the majority of the predictive power of treatment assignment in the future. Instruments showed a weakly sufficient conditional F-statistic, however autocorrelated errors may have skewed this. Certain confounding information was not available, and so we could only partly investigate balance across the instrument. We also could not empirically test the exclusion criteria.\par
Unstable estimates of the trajectory and highly collinear variables in the second stage model were the result, which were partly addressed using Ridge regression. However regularisation is unlikely to address the underlying problem of weak instruments, and more sophisticated techniques are needed going forward.\par
Further research may consider the use of techniques such as generalised least squares to account for heteroskedastic error terms in the models for the IV \cite{GLS}. Difficulties with highly correlated IVs could be attacked in one of two ways. Firstly, significantly improve the strength of the IV at time 1. Secondly, minimise the correlations between the instruments. Near far matching methodologies could identify pairs of GPs with the furthest possible difference in preference at initiation, or who are most different between two instrument times. We may also consider methods such as the Abrahamowics method, \cite{Bidulka} to attempt to model the point in which a physicians preference changes, to construct multiple IVs. \par
In a prior work \cite{Tompsett} prescription preference for Rheumatoid Arthritis treatment was taken over time to some success. However patents switched treatment strategies frequently, with a choice between more than two medications and the conditional F-statistic was dramatically higher than any identified for this study. This could indicate that time-varying IVs are more suited to studies with more than two treatment options .\par
To conclude, we have demonstrated and provided simple guidance to use R2SLS in a time varying instrument setting. The proposed method works well and is readily accessible to users in a variety of fields. Implementing a time-varying IV in practice is challenging due to high correlations between the IV over time. This makes it difficult to find sufficiently strong instrument strength which varies over time to estimate trajectories of causal effects. The viability of PP as a time varying instrument remains one of important research.

\subsection*{Acknowledgments}
The authors thank the Queen Mary University of London, Clinical Effectiveness Group and Professor Bianca De Stavola for the data.
\subsection*{Conflicts of Interest}
There are no conflicts of interest to declare.

\section*{Appendix}
\subsection*{Baseline Characteristics}
\begin{table}[H]
\begin{center}
\begin{tabular}{|l|l|l|}
\hline
Characteristics &\makecell{Sulfonylureas \\  (n=1733)} & \makecell{DPP4 \\ (n=828)}\\
\hline
Age (SD) & 56.8 (12.0) & 56.9 (11.7)\\
Male (\%)  & 53.8 & 48.5  \\
Ethnicity (\%) & & \\
\hspace{3mm}White  & 25.3 & 24.9 \\
\hspace{3mm}South Asian  & 50.4& 53.7 \\
\hspace{3mm}Black  & 19.1 & 16.5  \\
\hspace{3mm}Other & 5.1 & 4.8  \\
Non-Smoker (\%)  & 28.0 & 27.7  \\
\makecell{HbA1c level (SD) \\(mmol/mol)} & 67.7 (17.6) & 65.8 (15.5)  \\
\makecell{Median HbA1c Prior (SD) \\ to Initiation   \\(mmol/mol)} & 64.2 (17.4) & 61.5 (15.4)  \\
Body Mass Index (\%) & & \\
\hspace{3mm}$<25$ & 19.1 & 17.3 \\
\hspace{3mm}$25-29$  & 38.3& 29.9  \\
\hspace{3mm}$30-34$   & 25.4 & 25.5  \\
\hspace{3mm}$>34$& 17.2 & 27.1  \\
Years of Diabetes & & \\
\hspace{3mm}$<3$ yrs & 29.4 & 28.3  \\
\hspace{3mm}$3-5$ yrs  & 20.8 & 18.1  \\
\hspace{3mm}$>5$ yrs   & 49.7 & 53.5  \\
Systolic Blood Pressure & & \\
\hspace{3mm}$<100$  & 1.61 & 0.60  \\
\hspace{3mm}$100-139$   & 83.8 & 82.8  \\
\hspace{3mm}$>140$    & 14.5& 16.5  \\
Total Serum Cholesterol (mmol/L) (SD)& 4.12 (1.05) & 4.14 (1.10) \\
Glomerular Filtration Rate (eGFR) & & \\
\hspace{3mm}$>60$  &  91.6& 92.1 \\
\hspace{3mm}$30-59$   & 7.2 & 6.8\\
\hspace{3mm}$<30$    & 1.0& 0.9  \\
No of  Hypoglycemic events & & \\
\hspace{3mm}$0$  &  98.1& 98.1  \\
\hspace{3mm}$1+$  & 1.9 & 1.9 \\
Number of strokes  & & \\
\hspace{3mm}$0$  &  98.9& 99.2  \\
\hspace{3mm}$1$   & 1.1 & 0.8 \\
Medications (\%) & & \\
\hspace{3mm}Statins  &  86.3& 89.1 \\
\hspace{3mm}Beta-Blockers  & 16.6 & 17.0\\
\hline
\end{tabular}
\caption{Patient baseline characteristics at initiation of second line therapy.}
\end{center}
\end{table}
\subsection*{R2SLS Relation to g-estimation}
In g-estimation the substantive model of interest is a Structural Nested Mean Model (SNMM)
\begin{equation}
E[Y(\bar{a}_t,0)-Y(\bar{a}_{t-1},0)|\bar{A}_t=\bar{a}_t, \bar{Z}_t, M_t]= \beta_t a_t\qquad  t=1,\ldots,T-1.
\end{equation}
where $Y(\bar{a}_t,0)$ is the counterfactual outcome under a patient's observed treatment history up to time $t$, and on control afterwards.\par
Under the assumptions of no current treatment interaction, and a linear relationship bewteen $A$ and Y (see \cite{Tompsett}) the SNMM can be reduced to a Marginal Structural Model of the form
$$E[Y(a)]= \alpha+\sum_{t=1}^T \beta_t a_t.$$
Notably this model implies the Linear Structural Mean Model of 2SLS, which models the observed outcome Y as
$$E(Y|A,Z,M)=E[Y_0|A,Z,V]+\sum_{t=1}^T\beta_t A_t$$
In order to prove the equivalence of RS2LS and G-estimation with time varying IVS, we demonstrate the equivalence of their closed form solutions.\par
Define $$A=\begin{bmatrix}
A_{11}& \ldots & A_{T1}\\
\vdots & \vdots & \vdots \\
A_{1n} & \ldots & A_{Tn}\\
\end{bmatrix}$$
and define $Y$, $\hat{A}$, $Z$ equivalently. For notational convenience we label $Z^{res}=Z$
G estimation has the closed form solution. 
$$\beta^{gest}=Y^{'}ZZ^{'}A(A^{'}ZZ^{'}A)^{-1}$$
and the robust 2SLS method can be shown to have the closed form solution
$$\beta^{2SLS}=(A^{'}PA)^{-1}(A^{'}PY)$$
where P is the projection matrix 
$$P=Z(Z^{'}Z)^{-1}Z^{'}$$
and thus the transpose can be written as
$$ \beta^{2SLS'}=[(A^{'}PA)^{-1}(A^{'}PY)]^{'}$$
$$=(A^{'}PY)^{'}[(A^{'}PA)^{-1}]^{'}$$
$$=Y^{'}P^{'}A[(A^{'}PA)^{'}]^{-1}$$
$$=Y^{'}PA[(A^{'}PA)]^{-1}$$
Due to $P$ being symmetric.\par
We see that the two closed form solutions are similar with $P$ replacing $ZZ^{'}$, and are in fact equal as we have, expanding $P$
$$\beta^{2SLS'}=Y^{'}Z(Z^{'}Z)^{-1}Z^{'}A[(A^{'}Z(Z^{'}Z)^{-1}Z^{'}A)]^{-1}$$
noting that $Z^{'}A$, $(Z^{'}Z)^{-1}$ and $A^{'}Z$ are square matrices, we can rewrite the inverse term as 
$$=Y^{'}Z(Z^{'}Z)^{-1}Z^{'}A(Z^{'}A)^{-1}(Z^{'}Z)(A^{'}Z)^{-1}$$
$$=Y^{'}Z(A^{'}Z)^{-1}$$
$$=Y^{'}Z(Z^{'}A)(Z^{'}A)^{-1}(A^{'}Z)^{-1}$$
$$=Y^{'}ZZ^{'}A(A^{'}ZZ^{'}A)^{-1}=\beta^{gest}$$
Thus these two methods are equivalent, under the assumption that there are as many $Z$ as $A$, both the first and second stage models are fit by OLS, and that $Z^{'}A$, $(Z^{'}Z)^{-1}$ and $A^{'}Z$ are invertable matrices.

\subsection*{Penalised Regression Methods for Second Stage Model}
One challenge that may be encountered with time dependent instruments is that of collinearity. In situations in which a treatment does not vary much over time, the $A_t$ will be highly correlated. If the $Z_t$ also show only minor change over time, we are faced with the possibility that the majority of the predictive power of the $Z_t$ lies within the first time period. That is to say that each $A_t$ is far more strongly associated with $Z_1$ than any other $Z_t$.\par
Under these circumstances, the predicted $\hat{A}_t$ may be highly correlated and the second stage model may suffer from multicollinearity. This can lead to highly unstable estimates of the time specific causal effects $\beta_t$ and large standard errors.\par
To combat this we consider fitting the second stage model using Ridge regression methods \cite{Ridge,Ridge2}. Ridge regression fits the OLS model with penalty function determined by a tuning parameter $\lambda$. The solution to the second stage model is then.
\begin{equation}
\beta=min_{\beta}\{\sum_{i=1}^{n}(Y-\alpha-\beta_t\hat{A_t})^2 +\sum_{t=1}^{T-1}\lambda \beta_t^2 \}
\end{equation}
which is better known in its closed form solution 
$$\beta=(\hat{A}^{t}\hat{A}+\lambda I)^{-1}\hat{A}^{t}Y$$
with $I$ the identity matrix of appropriate dimension.
We see that the penalty $\lambda$ penalises large values of $\beta_t$ reducing instability in the estimates, and breaking the collinearity of the $\hat{A}_t$. The compromise is that it introduces bias to $\beta_t$. The larger the value of $\lambda$ the greater penalty to larger estimates of $\beta$, which biases estimates towards zero. Thus the choice of $\lambda$ is important. To minimise bias, we seek to choose the smallest $\lambda$ such that the colinearity in the second stage model is suitably broken to provide stable estimates of $\beta_t$. This is often done using ridge trace plots to observe the range in which estimates stabilise but can be formalised with the following algorithm.
\begin{enumerate}
\item Select a suitably wide range of values of $\lambda$, and take a set of $k$ increasing values $\lambda_k$ over this range
\item Perform 2SLS using Ridge regression in the second stage with each penalty $\lambda_k$. For each $k$, calculate the change in each parameter from the previous run. $\|\beta^{k}-\beta^{k-1}\|$
\item Select the first parameter $\beta^{k}$ for which this difference is within some chosen tolerance $tol$ for each $\beta_t$
\end{enumerate}
The tolerance should be small enough to attain stable estimates, but too small will lead to  choosing larger penalties and shrink estimates. Penalised regression methods have been considered in the first stage model in the context of Multivariate Mendelian randomisation to select from a high number of genetic instruments \cite{Robust}, but to our knowledge not in this context in the second stage model. We favour Ridge over other types of penalised regression because it cannot shrink estimates to 0.

\subsection*{Additional Simulations}
\subsubsection*{Probit based First Stage Models for Binary Data}
As with most theory on 2SLS, we have made the assumption that the first and second stage models are fit using Ordinary Least Squares (OLS). This creates a dilemma when the treatment and/or outcome are binary. Our approach is to fit an OLS model to the binary data, known as a linear probability model (LPM). LPMs however may not fit such data particularly well, where we might be expect the distribution near probabilities close to 0 or 1 to become non linear. \par 
One commonly explored avenue is to fit the first stage models using a probit or logistic regression model. These are often referred to as Two Stage Predictor Substitution (2SPS) \cite{Basu} or "forbidden" regression models \cite{Forbidden}. Whilst non OLS models may fit the binary data better, their consistency is still a matter of debate.\par
In the case of a binary treatment, non OLS first stage models cannot guarantee that the residuals of the model are uncorrelated with the predicted values $\hat{A}_t$. One noted trick however \cite{Forbidden} is as follows
\begin{enumerate}
\item Use 2SLS to obtain predictions of the treatment $\hat{A}_t$ from a first stage model with chosen non OLS link function.
\item Now perform 2SLS with an OLS first and second stage model, substituting the instruments $Z_t$ with the predictions $\hat{A}_t$  
\end{enumerate} 
This has been shown to be an efficient and consistent alternative to other 2SLS approaches \cite{Forbidden}.\par
In this paper we will only consider continuous outcomes. Two stage methods with a binary outcome are a greater challenge. Methods using a second stage LPM have been shown to be mainly unbiased, but most non LPM methods appear to be inconsistent \cite{Forbidden,Basu}, and depends on rarity of the outcome. This is beyond the current scope of the paper.
We repeat the simulations of the main body of the paper, whilst additionally testing R2SLS in the case of a probit first stage model, and using the "trick" method described above. In the simpler scenario, results show good or even slightly better performance that 2SLS or R2SLS with an OLS model, but show notable bias in the complex case. This indicates that the residualisation of the instruments is not effective when not applying OLS models.
\begin{table}[H]
\begin{center}
\begin{tabular}{|c|c|c|c|c|c|}
\hline
 n &  $\alpha$ & Bias & RMSE & MCE &Coverage\\
\hline
\multicolumn{6}{|c|}{Standard 2SLS}\\
\hline
5000 & 0.5 & 0.004 & 0.42& 0.006 & 93.5\\
5000 & 0.3 & 0.012 & 0.52& 0.009 & 93.6\\
5000 & 0.1 & 0.096 & 0.85& 0.022 & 94.2\\
1000 & 0.5 & 0.013 & 0.63& 0.012 & 94.5\\
1000 & 0.3 & 0.048 & 0.79& 0.020 & 95.6\\
1000 & 0.1 & 0.128 & 4.18& 0.554 & 98.9\\
\hline
\multicolumn{6}{|c|}{R2SLS}\\
\hline
5000 & 0.5 & 0.004 & 0.42& 0.006 & 93.5\\
5000 & 0.3 & 0.012 & 0.52& 0.009 & 93.6\\
5000 & 0.1 & 0.096 & 0.85& 0.022 & 94.2\\
1000 & 0.5 & 0.013 & 0.63& 0.012 & 94.5\\
1000 & 0.3 & 0.048 & 0.79& 0.020 & 95.6\\
1000 & 0.1 & 0.128 & 4.18& 0.554 & 98.9\\
\hline
\multicolumn{6}{|c|}{R2SLS (Probit)}\\
\hline
5000 & 0.5 & 0.027 & 0.42& 0.006 & 94.0\\
5000 & 0.3 & 0.022 & 0.52& 0.009 & 93.5\\
5000 & 0.1 & 0.054 & 0.84& 0.022 & 94.1\\
1000 & 0.5 & 0.032 & 0.62& 0.012 & 94.4\\
1000 & 0.3 & 0.040 & 0.78& 0.020 & 94.4\\
1000 & 0.1 & 0.400 & 1.96& 0.121 & 98.9\\
\hline
\multicolumn{6}{|c|}{R2SLS (Probit "Trick")}\\
\hline
5000 & 0.5 & 0.003 & 0.42& 0.006 & 93.5\\
5000 & 0.3 & 0.008 & 0.52& 0.009 & 93.6\\
5000 & 0.1 & 0.049 & 0.84& 0.022 & 94.1\\
1000 & 0.5 & 0.006 & 0.63& 0.012 & 94.2\\
1000 & 0.3 & 0.023 & 0.78& 0.020 & 95.1\\
1000 & 0.1 & 0.342 & 1.56& 0.076 & 99.9\\
\hline
\end{tabular}
\caption{Simulation results for Simple data setup of Figure 1, based on b=1000 bootstrapped samples. LPM and Probit indicate first stage models fit using OLS or a probit regression model respectively. Probit "Trick" denotes the method in which first stage predictions of $A_t$ using probit are subsequently used as instruments in 2SLS.  }
\end{center}
\end{table}

\begin{table}[H]
\begin{center}
\begin{tabular}{|c|c|c|c|c|c|}
\hline
 n &  $\alpha$ & Bias & RMSE & MCE &Coverage\\
\hline
\multicolumn{6}{|c|}{Standard 2SLS}\\
\hline
5000 & 0.5 & 1.900 & 1.38& 0.003 & 0.0\\
5000 & 0.3 & 2.513 & 2.81& 0.238 & 76.3\\
5000 & 0.1 & 3.397 & 1.91& 0.041 & 2.7\\
1000 & 0.5 & 1.895 & 1.38& 0.007 & 0.2\\
1000 & 0.3 & 2.781 & 4.12& 0.531 & 78.3\\
1000 & 0.1 & 3.361 & 2.87& 0.240 & 40.9\\
\hline
\multicolumn{6}{|c|}{R2SLS}\\
\hline
5000 & 0.5 & 0.010 & 0.33& 0.003 & 96.1\\
5000 & 0.3 & 0.014 & 0.42& 0.006 & 95.2\\
5000 & 0.1 & 0.638 & 3.21& 0.325 & 96.2\\
1000 & 0.5 & 0.006 & 0.50& 0.008& 95.8\\
1000 & 0.3 & 0.023 & 0.644& 0.013 & 95.4\\
1000 & 0.1 & 5.352 & 11.439& 4.140 & 98.7\\
\hline
\multicolumn{6}{|c|}{R2SLS (Probit)}\\
\hline
5000 & 0.5 & 0.535 & 0.74& 0.004 & 1.2\\
5000 & 0.3 & 0.220 & 0.54& 0.006 & 79.0\\
5000 & 0.1 & 0.527 & 1.44& 0.063 & 95.7\\
1000 & 0.5 & 0.505 & 0.76& 0.009 & 60.2\\
1000 & 0.3 & 0.169 & 0.70& 0.014 & 94.6\\
1000 & 0.1 & 0.331 & 3.36& 0.357 & 98.6\\
\hline
\multicolumn{6}{|c|}{R2SLS (Probit "Trick")}\\
\hline
5000 & 0.5 & 0.206 & 0.49& 0.004 & 59.2\\
5000 & 0.3 & 0.103 & 0.46& 0.006 & 91.4\\
5000 & 0.1 & 1.287 & 3.90& 0.479 & 98.1\\
1000 & 0.5 & 0.187 & 0.57& 0.008 & 90.6\\
1000 & 0.3 & 0.060 & 0.66& 0.013 & 95.5\\
1000 & 0.1 & 41.922 & 35.39& 39.603 & 99.6\\
\hline
\end{tabular}
\caption{Simulation results for Complex data setup of Figure 1, based on b=1000 bootstrapped samples. LPM and Probit indicate first stage models fit using OLS or a probit regression model respectively. Probit "Trick" denotes the method in which first stage predictions of $A_t$ using probit are subsequently used as instruments in 2SLS.}
\end{center}
\end{table}

\subsection*{Full Results: Case Study}

\begin{figure}[H]
\includegraphics[width=0.5\textwidth]{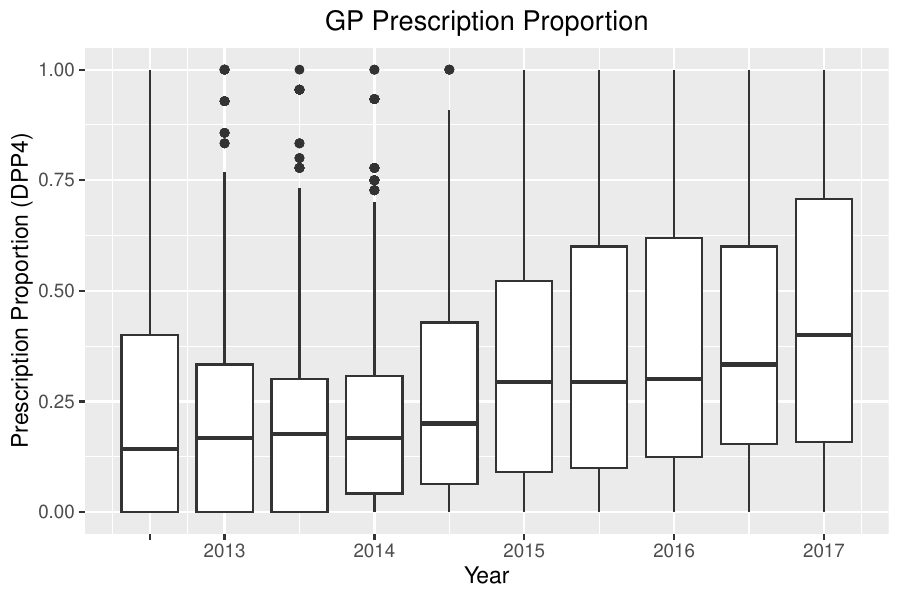}
\includegraphics[width=0.5\textwidth]{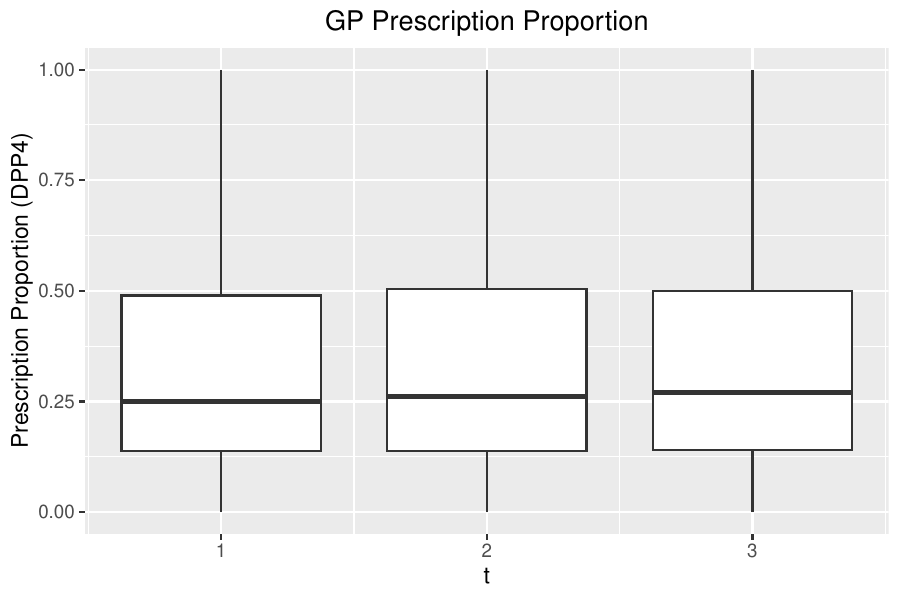}
\caption{Plot of trends of $PP^{Cal}$ and $PP^{t}$ over time.}
\end{figure}

\begin{table}[H]
\begin{center}
\begin{tabular}{|c|c|c|c|c|}
\hline
Model& \makecell{$\beta_1$\\ (95\% CI)} & \makecell{$\beta_2$\\ (95\% CI)} & \makecell{$\beta_3$\\ (95\% CI)} & \makecell{ATE\\ (95\% CI)}\\ 
\hline
\makecell{Standard 2SLS \\$PP^{Cal}$} & \makecell{-33.49 \\(-109.20,40.262)} & \makecell{35.82 \\ (-70.76,162.19)}  & \makecell{-7.08\\(-107.60,84.89)}  & \makecell{-4.75 \\(-7.11,-2.52)}  \\
\hline
\makecell{ R2SLS \\$PP^{Cal}$}  & \makecell{-9.09 \\(-51.97,36.11)} & \makecell{-21.00 \\ (-98.90,34.39)}  & \makecell{-27.61 \\(-106.90,83.74)}  & \makecell{-2.46 \\(-4.45,-0.74)}  \\
\hline
\makecell{Ridge R2SLS \\$PP^{Cal}$} & \makecell{-1.04 \\ (-1.63,-0.27)} & \makecell{-0.88 \\ (-1.57,-0.26)}  & \makecell{-0.56 \\ (-1.29,0.07)}  & \makecell{-2.49 \\ (-4.28,-0.65) }  \\
\hline
\makecell{Standard 2SLS \\$PP^{t}$} & \makecell{-22.14 \\(-62.24,8.31)} & \makecell{0.15 \\ (-55.88,58.43)}  & \makecell{16.64\\(-21.18,62.04)}  & \makecell{-5.34 \\(-8.29,-2.69)}  \\
\hline
\makecell{ R2SLS \\$PP^{t}$}& \makecell{-18.65\\ (-44.91,4.91)} & \makecell{18.39\\ (-16.18,55.71)}  & \makecell{-2.72\\ (-30.36,22.62)}  & \makecell{-2.99\\ (-5.47,-0.81)}  \\
\hline
\makecell{Ridge R2SLS \\$PP^{t}$}  & \makecell{-1.17\\ (-1.86,-0.36)} & \makecell{-0.72\\ (-1.53,0.01)}  & \makecell{-0.76\\ (-1.56,-0.02)}  & \makecell{-2.65\\ (-4.79,-0.66)}  \\
\hline 
\end{tabular}
\caption{Full Table of Results for Effect of DPP4 versus SU on HbA1c levels.}
\end{center}
\end{table}

\begin{table}[H]
\begin{center}
\begin{tabular}{|l|l|}
\hline
Characteristics & Correlations\\
\hline
Time period ($t$) & 0.174* \\
Age & 0.002 \\
Gender  & 0.012\\
Ethnicity & \\
\hspace{3mm}White  & -0.065 \\
\hspace{3mm}South Asian  & 0.094*\\
\hspace{3mm}Black  & -0.043\\
\hspace{3mm}Other & -0.008  \\
Smoke  & -0.082*   \\
HbA1c level (SD)  & -0.084*  \\
Median HbA1c  & -0.092*\\
Body Mass Index &  \\
\hspace{3mm}$<25$ & 0.003 \\
\hspace{3mm}$25-29$  & 0.026  \\
\hspace{3mm}$30-34$   & -0.032   \\
\hspace{3mm}$>34$& 0.022  \\
Years of Diabetes &  \\
\hspace{3mm}$<3$ yrs & 0.029   \\
\hspace{3mm}$3-5$ yrs  & -0.036   \\
\hspace{3mm}$>5$ yrs   & 0.000 \\
Systolic Blood Pressure & \\
\hspace{3mm}$<100$  & -0.045  \\
\hspace{3mm}$100-139$   & -0.031   \\
\hspace{3mm}$>140$    & 0.046\\
Total Serum Cholesterol (mmol/L) (SD)& 0.000 \\
Glomerular Filtration Rate (eGFR) &  \\
\hspace{3mm}$>60$  &  0.016 \\
\hspace{3mm}$30-59$   & -0.025 \\
\hspace{3mm}$<30$    & 0.021 \\
No of  Hypoglycemic events & 0.007 \\
Number of strokes  & -0.038 \\
Medications (\%) &  \\
\hspace{3mm}Statins  &  0.025 \\
\hspace{3mm}Beta-Blockers  & 0.012\\
\hline
\end{tabular}
\caption{Balance Table of correlations between confounders for $PP_1^{Cal}$. Variables with correlations above $0.08$ are marked, indicating potential imbalance.}
\end{center}
\end{table}

\bibliographystyle{ama}
\bibliography{refpaper4}

\end{document}